\documentclass[11pt,twoside]{article}

\usepackage{asp2006}
\usepackage{epsf}
\usepackage{psfig}
\usepackage{lscape}

\begin{document}
\title{A synthetic view of AGN evolution and supermassive black holes growth}   %%% Fill in title
\author{A. Merloni}   %%% Fill in author names
\affil{Excellence Cluster Universe, TUM, Boltzmannstr. 2, 85478, Garching
  \& \\ MPE, Giessenbachstr., Garching, Germany}    %%% Fill in author affiliations

\begin{abstract} %%% Abstract to run on from here.

I will describe the constraints available from a study of AGN evolution
synthesis models on the growth of the supermassive black holes (SMBH) 
population in the two main
'modes' observed (kinetic- and radiatively-dominated, respectively).
I'll show how SMBH mass function evolves anti-hierarchically,
i.e. the most massive holes grew earlier and faster than less massive
ones, and I will also derive tight  constraints on the
average radiative efficiency of AGN.
An outlook on the redshift evolution of the AGN kinetic
luminosity function will also be discussed, thus providing a robust
physical framework for phenomenological models of AGN
feedback within structure formation.
Finally, I will present new constraints on the evolution of
the black hole-galaxy scaling relation at $1<z<2$ derived by exploiting
the full multi-wavelength coverage of the COSMOS survey on a complete
sample of ~90 type 1 AGN.
\end{abstract}

%%% MAIN BODY OF TEXT GOES HERE. CONSULT "INSTRUCTIONS FOR AUTHORS USING
%%% LATEX2E MARKUP", SECTIONS 2.3-2.6 FOR HELP WITH EQUATIONS, FIGURES,
%%% AND TABLES.

%\section{}   %%% Top level section head (remove "%" symbol)
%\subsection{}   %%% Second level section head (remove "%" symbol)
%\subsubsection{}   %%% Lowest level section head (remove "%" symbol)
%\section*{}    %%% Unnumbered top level section head (remove "%" symbol)
%\subsection*{}   %%% Unnumbered second level section head (remove "%" symbol)

\section{Introduction}

In the past decade three seminal discoveries have revealed tight links
and feedback loops between the growth of nuclear super-massive black
holes and galaxy evolution, promoting a true
 shift of paradigm in our view of astrophysical black holes, which
have moved from the role of exotic tracers
of cosmic structures to
that of fundamental ingredients of them.

First of all, the search for the local QSO relics via the study of
their dynamical influence on the surrounding stars and gas
 led to the discovery of
SMBH in the center of most nearby bulge-dominated galaxies. The steep
and tight correlations between their masses and bulge properties
\citep[so-called {\it scaling
  relations};][]{gebhardt:00,ferrarese:00,haering:04} 
represented the first and
fundamental piece of evidence in favor of a connection between galaxy
evolution and central black holes.
The second one stems from the fact that SMBH growth is now known to be
due to radiatively efficient accretion over cosmological times, taking
place during ``active'' phases  \citep[hereafter
MH08]{marconi:04,merloni:08}. 
If most galaxies host a SMBH today, they should have
experienced such a phase of strong nuclear activity in the past.
Finally, extensive programs of optical and NIR follow-up observations
of X-ray selected AGN in the {\it Chandra} and {\it XMM-Newton} era
 put on solid grounds the evolution of accretion
luminosity over a significant fraction of cosmic time. We have thus
discovered that lower luminosity AGN peak at a lower
redshift than luminous QSOs \citep[see e.g.][]{hasinger:05}. Such a
 behavior is analogous to that observed for star
formation (usually referred to as ``cosmic downsizing'') lending
further support to the idea that the formation and
evolution of SMBHs and their host galaxies might be closely related.

In this brief review, I will try to summarize observational evidences
for AGN downsizing, on the basis of a simple theoretical framework
according to which supermassive 
black holes evolution is dominated by accretion, and 
governed by a continuity equation, that we can solve numerically
between $z=0$ and $z\sim 4$. I will then show the implication for these
specific trends for the AGN energy release in kinetic form. Finally, I
will present some recent observational results on the evolution of the
scaling relations, and briefly discuss their implication for feedback
models.

\section{Dissecting AGN downsizing}

The term {\it downsizing} was first used by Cowie et al. (1996) to
describe their finding that actively star-forming galaxies at low
redshift have smaller masses than actively star-forming galaxies at
$z\sim 1$. In the
current cosmology jargon, this term has come to identify a
variety of possibly distinct phenomena, not just
related to the epoch of star formation, but also to that of star
formation quenching, or galaxy assembly (see the discussion in
Faber et al. 2007,
and references therein).  Given the growing body of observational
 evidence for galaxy downsizing, 
it is legitimate to ask whether black holes and AGN do also
 show a similar trend. The first hints of a positive answer came from
 the study of the evolution of the X-ray selected AGN luminosity
 function. 
In the last decade, we have learned 
that more luminous AGN were more common in
 the past, with the X-ray luminosity function (XLF) following a
 so-called Luminosity Dependent Density Evolution \citep{ueda:03,hasinger:05},
 a direct phenomenological manifestation of AGN
downsizing. 
How can we use this (and other analogous) results on the XLF evolution
to gain further insights on the physical evolution of the black hole population?

As opposed to the case of galaxies, where the direct relationship
between the evolving mass functions of the various morphological types
and the distribution of star forming galaxies is not straightforward
due to the never-ending morphological and photometric transformation
of the different populations, the situation in the case of SMBH is
much simpler. For the latter case, we can assume their evolution is
governed by a continuity equation (MH08, and references therein), 
where the mass function of SMBH at any given time can be used
to predict that at any other time, provided the distribution of
accretion rates as a function of black hole mass is known. 
Such equation  can be written as:
\begin{equation}
\label{eq:continuity}
\frac{\partial \psi(\mu,t)}{\partial t} +
\frac{\partial}{\partial \mu}\left( \psi(\mu,t) \langle \dot M
  (\mu,t)\rangle \right)=0
\end{equation}
where  $\mu=Log\, M$ ($M$ is the black hole mass in solar units),
$\psi(\mu,t)$ is the SMBH mass function at time $t$, and $\langle \dot M
(\mu,t) \rangle$ is the average accretion rate of SMBH of mass $M$ at
time $t$, and can be defined through a ``fueling'' function,
$F(\dot\mu,\mu,t)$, describing the
distribution of accretion rates for objects of mass $M$ at time $t$:
$\langle \dot M(M,z)\rangle = \int \dot M F(\dot\mu,\mu,z)\, \mathrm{d}\dot\mu$.
Such a fueling function is not a priori known, and observational
determinations thereof have been able so far to probe robustly only
the extremes of the overall population.   
However, the AGN fueling function 
can be derived by inverting the integral
equation that relates the luminosity function of the population in
question with its mass function. Indeed we can write:
\begin{equation}
\label{eq:filter}
\phi(\ell,t)=\int F(\ell-\zeta,\mu,t) \psi(\mu,t)\; \mathrm{d}\mu
\end{equation}
where I have called $\ell=Log\, L_{\rm bol}$ and $\zeta=Log\,
(\epsilon_{\rm rad}c^2)$, with $\epsilon_{\rm rad}$ the radiative
efficiency, here assumed to be constant.
This is the approach followed in MH08, were the inversion was
performed numerically,
based on a minimization scheme that used both the X-ray and radio AGN
luminosity functions as constraints, complemented by recipes to relate
observed (and intrinsic) X-ray and radio (core) luminosities to
$L_{\rm bol}$ (see MH08 for details).

\begin{figure}
\begin{tabular}{ll}
\plottwo{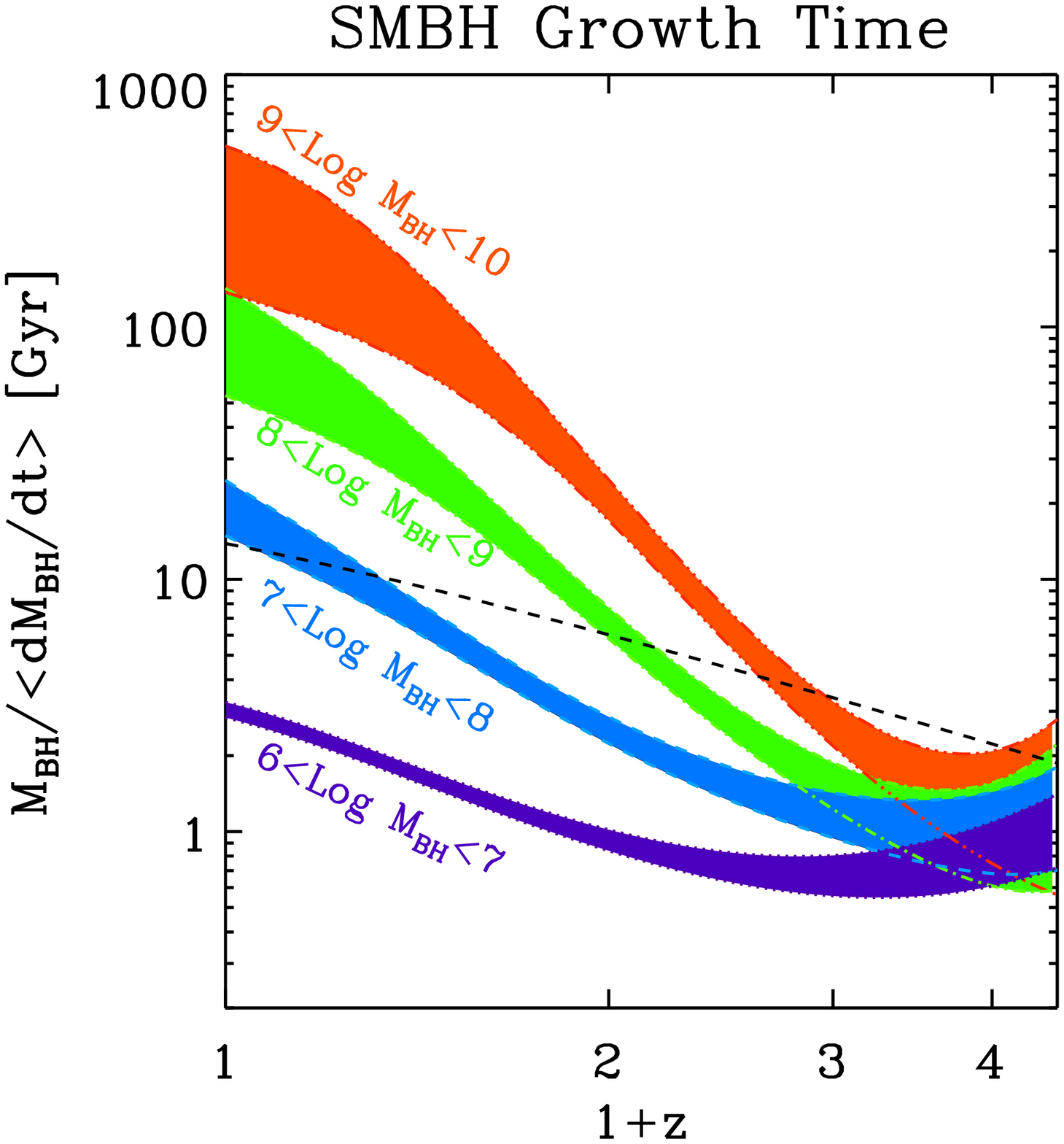}{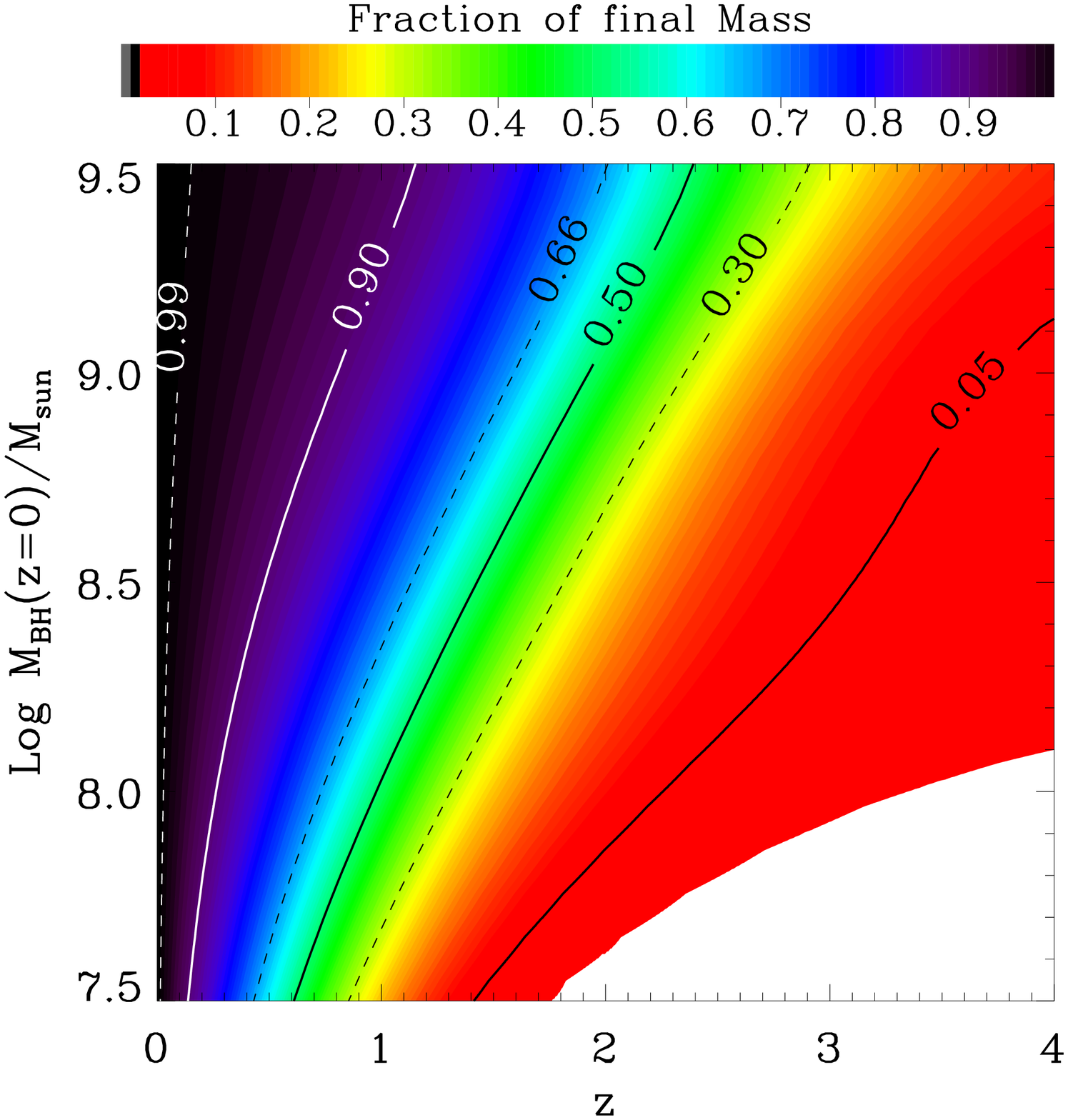}
\end{tabular}
  \caption{{\bf Left}:  Average Growth time of Supermassive Black Holes (in years)
    as a function of redshift for different black hole mass ranges. 
    The dashed line marks the age of the Universe; 
    only black holes with instantaneous
    growth time smaller than the age of the Universe at any particular
    redshift can be said to be effectively growing. {\bf Right}: the
    fraction of the final black hole mass accumulated as a function of
  redshift and final (i.e. at $z=0$) mass is plotted as contours.}
   \label{fig:xmm}
\end{figure}

Using this approach, 
we have integrated eq~(\ref{eq:continuity}) starting from $z=0$, where
we have simultaneous knowledge of both mass, $\psi(\mu)$, and
luminosity, $\phi(\ell)$, functions,
evolving the SMBH mass function backwards in time, up to where
reliable estimates of the 
AGN luminosity functions are available (currently this means $z\simeq 4$).
The adopted hard X-ray luminosity function is
supplemented with luminosity-dependent bolometric
corrections of Marconi et al. (2004)
and absorbing column density distributions 
consistent with the X-ray background  constraints, following
the most recent XRB synthesis model \citep{gilli:07}. 
Similar results can of course be obtained using directly
bolometric luminosity functions \citep[see e.g.][]{hopkins:07b}.

In this way, we can estimate the specific
instantaneous ratio of black hole mass to accretion rate as a function
of SMBH mass and its cosmological evolution. 
Such a ratio defines a timescale, the so-called {\it growth
time}, or mass doubling time (Figure~\ref{fig:xmm}, left). 
 The redshift evolution of the growth time distribution
can be used to identify the epochs when black holes of different sizes
grew the largest fraction of their mass: black holes
with growth times longer than the age of the Universe are not
experiencing a major growth phase, which must have necessarily
happened at earlier times. Figure~\ref{fig:xmm} then shows that, while
at $z \la 1$ only black holes with masses smaller than $10^7
M_{\odot}$ are experiencing significant growth, as
we approach the peak of the black hole accretion rate density ($z \sim
1.5-2$), we witness the rapid growth of the entire SMBH
population. Better constraints on both bolometric luminosity and mass
functions evolution are however needed to paint a clearer picture at
higher $z$. A strikingly similar behaviour has been indeed observed the
the whole of the star-forming galaxy population \citep{perez:08}.

Solutions of the continuity equation allow also to trace the
growth of black
holes of a given final (i.e. at $z=0$) mass. The right hand side panel of
Fig.~\ref{fig:xmm} shows that, for the most massive black holes
($>10^9 M_{\odot}$) half of the mass was already in place at $z\sim
2$, while those with $M(z=0)<10^8 M_{\odot}$ had to wait until $z\sim
1$ to accumulate the same fraction of their final mass.

\section{Kinetic energy output of AGN}
The phenomenological investigation presented here, however, leaves
open the fundamental question about the physical origin of such a
clear, parallel differential growth of both the black holes and the
galaxy population. Some clues may come from attempts to trace the
evolution of the feedback energy released by growing black holes as a
function of black hole mass and redshift. In this section, we
describe in some detail how such an inventory can be made.

Direct evidence of AGN feedback in action has been found in the X-ray
observations of galaxy clusters, showing how black holes may deposit
large amounts of energy into their environment. From studies of the
cavities, bubbles and weak shocks generated by the radio emitting jets
in the intra-cluster medium (ICM) it appears that AGN are
energetically able to balance radiative losses from the ICM in the
majority of cases \citep{best:06,rafferty:06}.

\begin{figure}
  \plottwo{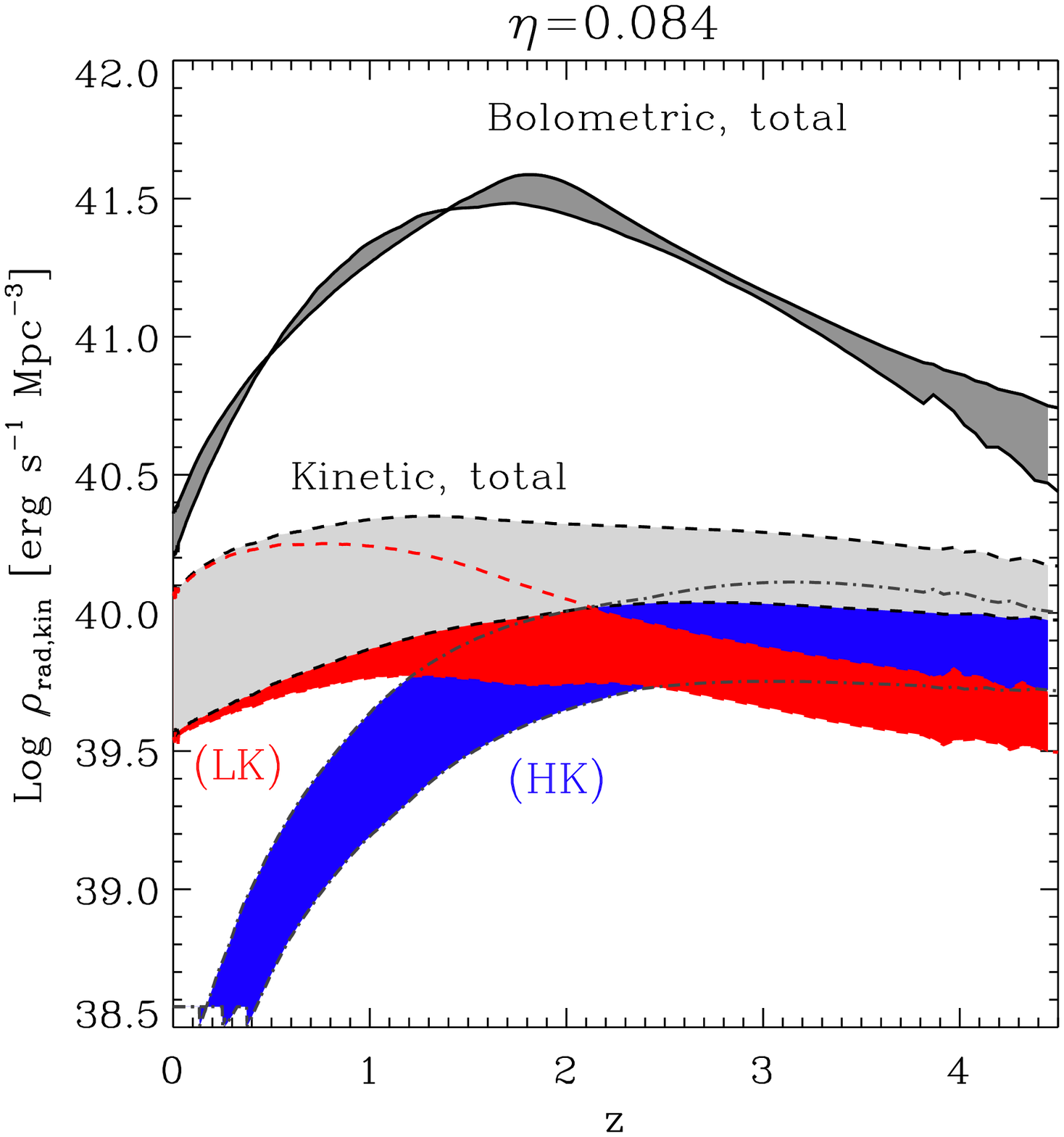}{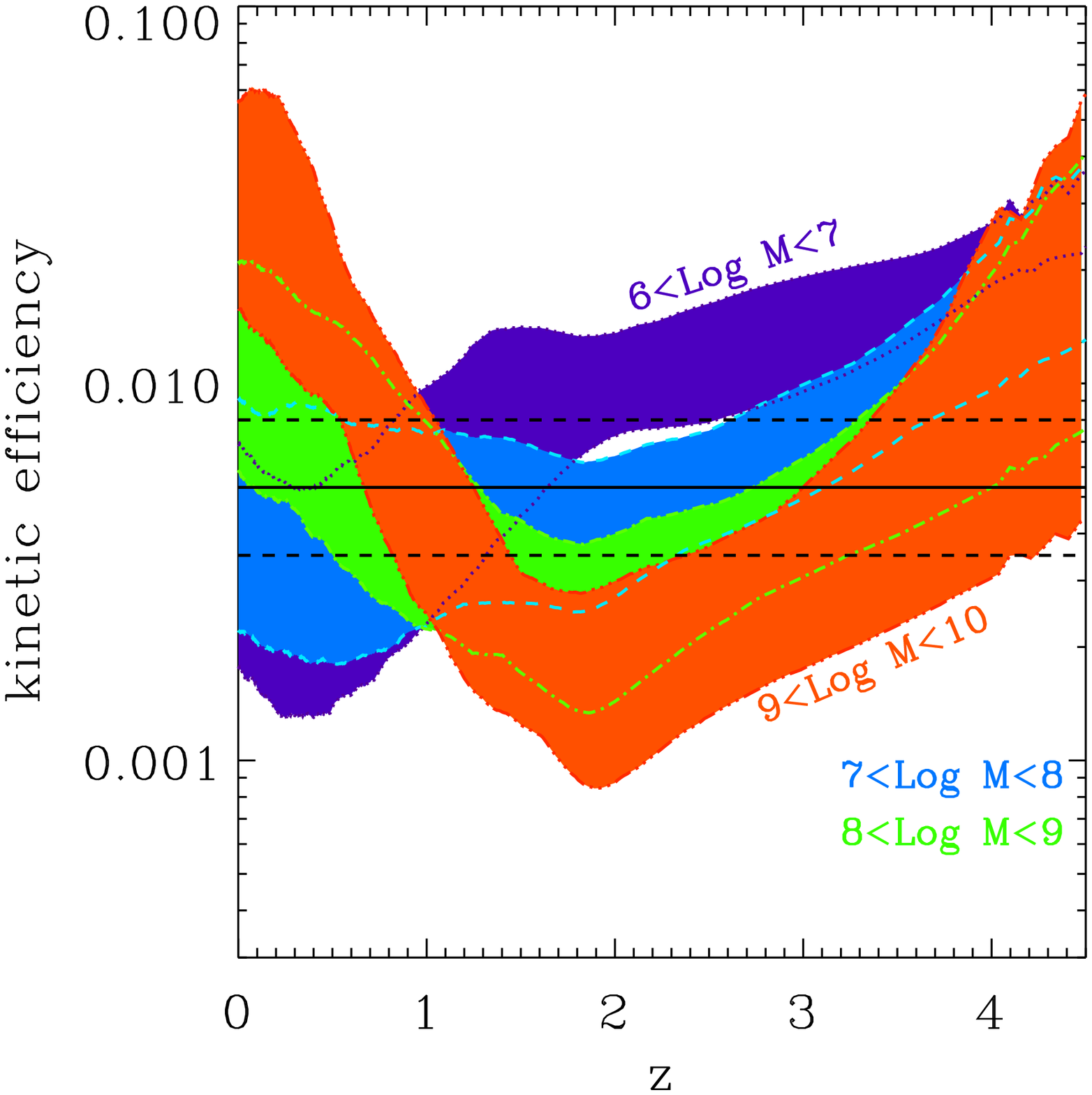}
  \caption{{\bf Left:} Redshift evolution of the radiative (bolometric) energy
  density of AGN compared to the estimated total kinetic one. The
  Kinetic energy density has been split into contribution from low and high
  accretion rate object (LK and HK, respectively). {\bf Right:} Redshift evolution of the kinetic efficiency $\epsilon_{\rm
    kin}$. SMBH of different masses are here plotted separately.
The horizontal black solid line marks the mass weighted average
values for the kinetic efficiency, with the dashed lines
representing the uncertainties. In both plots, shaded areas
represents the uncertainty 
deriving from the observational uncertainty on the AGN
luminosity functions.}
   \label{fig:kin_eff}
\end{figure}

On the other hand, numerical simulations of AGN-induced feedback have
shown that mechanical feedback from black holes may be responsible for
halting star formation in massive ellipticals, explaining the
bimodality in the color distribution of local galaxies
\citep{springel:05}, 
as well as the size of the most massive ones. At a global
level, these models hinge on the unknown efficiency with which growing
black holes convert accreted rest mass into kinetic and/or radiative
power. Constraints on these efficiency factors are therefore vital for
the theory.
Recent works of ours \citep{merloni:07} showed that the output of low-luminosity AGN is
dominated by kinetic energy rather than by radiation and have allowed
estimates of the kinetic luminosity function of AGN based on the
observed radio emission of their jets 
(either core, MH08, or extended, Cattaneo \& Best 2009). 

 As opposed to the
mass growth evolution, the kinetic luminosity function so derived is
not very tightly constrained due to poor observational information on the true
(intrinsic) radio core/extended luminosity functions and to the large
uncertainties in the calibration of the empirical relations between
total kinetic power and radio luminosity. Nevertheless, we can constrain
the local ($z=0$) AGN kinetic power density, $\rho_{\rm kin}$, 
between 1 and 10 $\times$ 10$^{39}$ erg s$^{-1}$ Mpc$^{-3}$, comparable with
the total kinetic power density from type II Supernovae ($\rho_{\rm
  SNII}\simeq 4 \times 10^{39}$  erg s$^{-1}$ Mpc$^{-3}$,
Hopkins \& Beacom 2006), 
with the total AGN radiative density being about
$\rho_{\rm rad}(z=0)\simeq 1.6 \times 10^{40}$ erg s$^{-1}$ Mpc$^{-3}$.

Integrating over redshift, we are able to measure the overall
efficiency of SMBH in converting accreted rest mass energy into
kinetic power, the  ``kinetic efficiency'' $\epsilon_{\rm kin} \equiv
L_{\rm kin}/(\dot M c^2)$,
 which ranges between 3 and 8 $\times 10^{-3}$ (MH08), depending on
the choice of the radio core luminosity function, or between 1 and 10
$\times 10^{-3}$ \citep{cattaneo:09}, depending on the exact choice of
the $L_{\rm kin}$-$L_{\rm radio}$ relation. This is to be compared to
the radiative efficiency, $\epsilon_{\rm rad}$, approximately 
constrained to be between 0.07 and 0.16, depending on the choice of
the local BH mass function, bolometric and obscuration corrections
\citep[see e.g.][]{marconi:04,merloni:08,yu:08}. As mentioned before,
this implies that the overall growth of SMBH happens ``radiatively'',
with mechanical power output representing a small, albeit significant,
fraction of the energy release.

The possibility to resolve the mass and accretion rate distribution
functions with the continuity equation approach 
allows us to separate the evolution of both growth rate and kinetic
energy density into different mass bins and into the various modes of
accretion. We found that the kinetic power density at low redshift is
dominated by low luminosity AGN, while the contribution
from radio loud QSOs becomes significant at $z\sim2$
(see left panel in Fig.~\ref{fig:kin_eff}).  The measured $\epsilon_{\rm kin}$
varies strongly with SMBH mass and redshift, being maximal for very
massive holes at late times, a property in agreement with what
required for the AGN feedback by many recent galaxy formation models
(right panel of Fig.~\ref{fig:kin_eff}).

\section{The evolution of scaling relations}

In the two previous sections, we have described methods to extract
information from observations at various wavelengths about the
differential growth of supermassive black holes and the corresponding
mechanical energy output. However, direct evidence that AGN feedback
may be responsible for the observed downsizing of galaxies is still
missing. Observational clues could be hidden in the cosmological evolution of the
scaling relations between SMBH and hosts, and, indeed, modern multiwavelength
surveys are increasingly designed to allow measurements of the
physical properties of AGN hosts. 

\begin{figure}
\plotfiddle{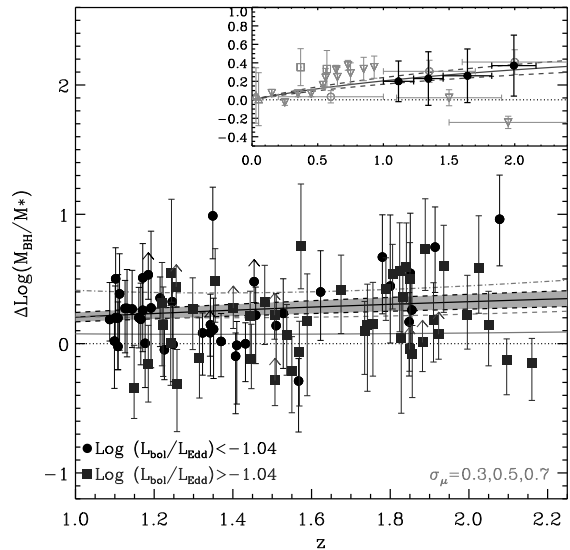}{6cm}{0}{33}{33}{-110}{-30}
%\plotone{dmbh_z_mass_all_bw.ps}
  \caption{{\bf Left:} Redshift evolution of the offset measured for zCOSMOS type--1 AGN from
  the local $M_{\rm BH}-M_{*}$ relation. Different symbols identify different
ranges of Eddington ratios with upwards arrows representing upper limits on 
the host mass. The solid black line and the grey shaded area
show the best fit obtained assuming an evolution of the form
$\Delta {\rm Log}(M_{\rm BH}/M_{*})(z)=\delta_2{\rm Log}(1+z)$; for which we
found $\delta_2=0.68 \pm 0.12$. Grey lines show the bias due to the intrinsic
scatter in the scaling relation to be expected even if they are universal. Solid line
is for an intrinsic scatter of 0.3 dex; dashed of 0.5 dex; dot-dashed
of 0.7 dex. The inset shows a comparison with literature results. 
%{\bf
%  Right:} Black hole mass host stellar mass relation. 
%Symbols with leftwards arrows represent upper limits on
%the host mass. The black solid line is the best fit to the
%\cite{haering:04} local spheroids sample relation, with dashed lines
%marking a $\pm$0.3 dex offset.  Green/grey
%arrows represent the direction of evolution of the points in the
%$M_{\rm BH}$-$M_{*}$ plane in 300 Myr on the basis of their instantaneous
%accretion- and star formation-rates and an AGN duty-cycle estimated
%from the amplitude of the corresponding luminosity and mass functions.
%From Merloni et al. (2009).}
}
   \label{fig:dmbh_z}
\end{figure}

Within the COSMOS survey \citep{scoville:07}, we have recently studied the hosts of 89 broad line (type--1) Active Galactic Nuclei (AGN)  detected
in the zCOSMOS survey in the
redshift range $1<z<2.2$ (for all the details, see Merloni et al. 2009). 
The unprecedented multi-wavelength coverage of
the survey field allowed us to
disentangle the emission of the host galaxy from that of the nuclear
black hole in their Spectral Energy Distributions (SED).
We derive an estimate of black hole masses
through the analysis of the broad MgII emission
lines observed in the
medium-resolution spectra taken with {\it VIMOS/VLT} as part of the
zCOSMOS project. Then, we estimated rest frame K-band luminosity and
total stellar 
mass (and their corresponding uncertainties) of the AGN hosts through
an extensive SED fitting procedure, based on large databases of both
phenomenological and theoretical galaxy spectra.

We found that, as compared to the local value, the
average black hole to host galaxy mass ratio appears to evolve positively with
redshift, with a best fit evolution of the form
$(1+z)^{0.68 \pm 0.12 ^{+0.6}_{-0.3}}$ (see Fig.~\ref{fig:dmbh_z}), 
where the large asymmetric systematic
errors stem from the uncertainties in the choice of IMF, in the
calibration of the virial relation used to estimate BH masses and in
the mean QSO SED adopted. 
A thorough analysis of observational biases induced by
intrinsic scatter in the scaling relations reinforces the conclusion
that an evolution of the $M_{\rm BH}-M_{*}$ relation must
ensue for actively growing black holes at early times:
either its overall normalization, or its intrinsic scatter (or
both) appear to increase with redshift.

\subsection{Implications for theoretical models}
Such an evolution is at odds with the predictions of essentially 
all feedback models in which the
black hole energy injection is very fast (explosive). Indeed, the
first published predictions of merger-induced AGN activity models
\citep{robertson:06} indicated that, if strong QSO feedback is
responsible for rapidly terminating star formation in the bulge, as in
the models of Di Matteo et al. (2005), then very little evolution, as well
as very little scatter, is expected for the scaling relations.
However, later works within the same theoretical framework 
\citep{hopkins:09a}
have analyzed in greater depths the role of dissipation in major
mergers at different redshift. Under the assumption that black hole
and spheroids obey a universal ``black hole fundamental plane'', 
where $M_{\rm BH}\propto M_{*}\sigma_{*}^2$, they show
how, in
gas-richer environments (at higher redshift), dissipation effects may
deepen the potential well around the black hole, allowing it to grow
above the local $M_{\rm BH}-M_{*}$ relation, to a degree marginally
consistent with our results. 
Another, related effect was discussed in Croton (2006). There it was
assumed that major mergers can trigger both star formation in a bulge
as well as black hole growth, in a fixed proportion. However,
bulges can also acquire mass by disrupting stellar discs, a channel
that should not contribute to black hole growth. The relative
importance of these two paths of bulge formation may lead to lighter
bulges for a given black hole mass at high redshift, as disks have a
smaller stellar fraction (see also Malbon et
al. 2007).

\section{Concluding remarks}
The overall growth of SMBH through accretion is now quite well
sampled, mainly thanks to multi-wavelength coverage of X-ray selected
AGN in large surveys. The main missing ingredients here are the exact
census of the heavily obscured (Compton thick) objects and a more
robust determination of bolometric corrections as a function of
luminosity, black hole mass and Eddington ratios \citep{vasudevan:07}.
The kinetic luminosity function of AGN is less well constrained
than the radiative one, due to a poorer knowledge of both high
redshift radio luminosity functions and the robust calibration of the
relationship between mechanical power and radio luminosity.

Notwithstanding these last open issues, we have argued here that
strong observational indications can be gathered on the cosmological
evolution of the SMBH population and of its overall energy output
rate. The final link between such studies and the understanding of the
physical relations between growing black holes and their hosts 
will rest on future progresses in the joint study of AGN and galaxies.
In particular, the redshift evolution of the scaling relations will
likely be an important testbed for structure formation models.

From the observational point of view, it will be very important to
explore methods to derive robust black hole mass estimates in high
redshift samples of obscured AGN, that can be 
selected purely on the basis of their host galaxy
properties. Broad emission lines at longer wavelengths, where the
effect of obscuration are less severe, could be very useful in this
respect. Also, a better understanding of the differences in the
hosts' properties of active and inactive black holes is needed to
allow a more meaningful comparison with the local scaling relations,
and a better assessment of their evolution.
From the theoretical point of view, more efforts should be devoted to
derive robust predictions for the coupled
evolution of slope, normalization {\it and intrinsic scatter} in the scaling
relations.
%%%%%%%%%%%%%%%%%%%%%%%%%%%%%%%%%%%%%%%%%%%%%%%%
%% BACKMATTER
%%%%%%%%%%%%%%%%%%%%%%%%%%%%%%%%%%%%%%%%%%%%%%%%

\acknowledgments
I am grateful to my collaborators A. Bongiorno, M. Brusa, S. Heinz,
and all the members of the COSMOS and zCOSMOS teams for their 
essential contribution to the work presented here.

\end{document}